# Analytical Heat Transfer Modeling of a New Radiation Calorimeter

Elysée Obame Ndong[1,2], Olivier Gallot-Lavallée[2], Frédéric Aitken[2]
1. Department of Industry Engineering and Maintenance, University of Sciences and Technology of Masuku (USTM), BP 941, Franceville, Gabon
2. Grenoble Electrical Engineering Laboratory (G2Elab), Univ. Grenoble Alpes and CNRS, G2Elab, F38000, Grenoble, France

*Abstract*— **This paper deals with an analytical modeling of heat transfers simulating a new radiation calorimeter operating in a temperature range from -50 °C to 150 °C. The aim of this modeling is the evaluation of the feasibility and performance of the calorimeter by assessing the measurement of power losses of some electrical devices by radiation, the influence of the geometry and materials. Finally a theoretical sensibility of the new apparatus is estimated at ±1 mW. From these results the calorimeter has been successfully implemented and patented.**

## I. Introduction

The calorimetric method is one of the most reliable techniques for measuring heat power loss in electrical component [1]. Therefore several calorimetric devices have been developed for that purpose. They exhibit good measurements accuracies in their operating ranges. However, the investigations led on these calorimeters [2] show that these experimental setups point out a number of limits either on their temperature range, frequency, applied voltage or the component geometry. Seen these limits, we decided to design a new radiation calorimeter enabling to take into account the advantages presented by current calorimeters and improving their limits [3]. Our calorimeter has been designed for measuring heat power loss in a controlled insulated vacuum environment in which the operating temperature ranges from -50 °C to +150 °C. The tested electrical device can have any shape but this one cannot exceed a sphere diameter of 18 cm. The maximum measurable power is of the order of 10 W for an operating temperature of 100 °C and the precision of the measurement is of the order of 2% and less for heat power losses above 100 mW.

The design of our calorimeter apparatus was achieved from a thermal modeling which is the aim of this paper. The thermal modeling of our calorimeter is based mainly on an analytical approach because it allows varying easily many parameters in order to understand their effects and the amplitude of these effects on the performance of our device. The aim of this modeling is not to have a perfect true description of the thermal response of our device but to help to evaluate the performance of our device by choosing the best materials for the different part of our calorimeter. Also the objective of this modeling is to minimize the parasitic thermal losses such that the radiation heat transfer must be the dominating mode of heat transfer in our device. Our approach should account for the behavior both in steady-state and transient regimes.

The analytical approach requires simplifications of the heat exchange occurring in the calorimeter therefore it is not expected to have a fully quantitative description of our device. However, we will show that a good qualitative agreement has been achieved. At lower temperatures than room temperature a good quantitative agreement between our modeling and experiment is shown in the last chapter validating our approach.

## II. Brief presentation of the new calorimeter

Architecture for the calorimeter is shown on Fig. 1: it is composed by insulation, an isothermal system in which the electrical component to test is suspended, a heat sink and a heat vector conducting the heat power towards the heat sink. A more detailed description of the calorimeter architecture and geometry of its components have been given in Obame *et al.* [2] [3].

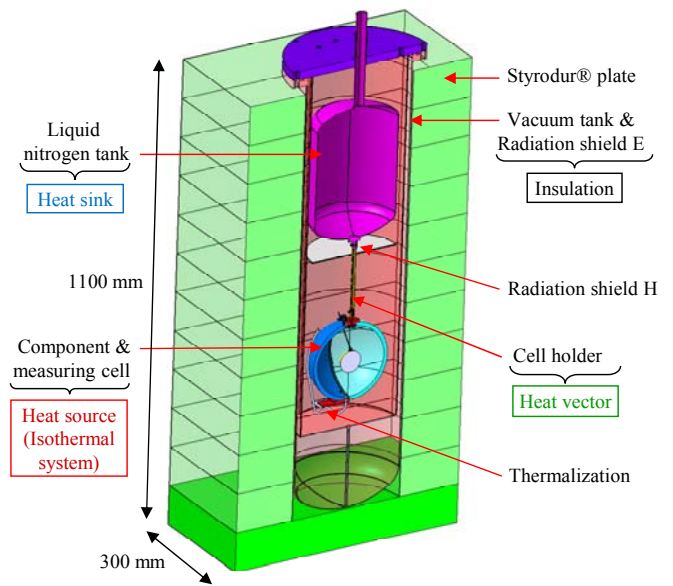

Fig. 1. Calorimetric longitudinal cross section with main functions of elements inside the system.

The principle used for measuring the heat power loss of an electrical device is a differential measurement method. The heat power dissipated by the electrical device is measured in two steps. For the first step, the electrical device under test is not supplied and the measuring cell is regulated at a given temperature $T_0$ by a set-up that provides heat power $P_0$ in steady state regime. In a second step, the electrical device is supplied and dissipates heat loss, mainly by radiation in our case. This loss disturbs the equilibrium temperature of the isothermal measuring cell and tends to increase it. The temperature controller reacts by lowering the heat supplied to the cell at a value $P_1$ in order to keep its temperature at the initial value $T_0$. Therefore, the heat loss dissipated by the electrical device can be calculated by $P_{loss} = P_0 - P_1$, assuming that the external disturbance of the cell can be neglected.

III. HEAT TRANSFER MODELING OF THE NEW CALORIMETER

The calorimeter being under vacuum, heat exchange by convection will occur only for external surfaces in contact with ambient air at room temperature. Then the analytical modeling developed here is mainly based on heat conduction and radiation.

For the need of the analytical approach the calorimeter has been divided in a great number of isothermal elements. All these elements have been indexed as shown on Fig. 2.

*A. Assumptions of the model*

In order to account for possible thermal gradients, the calorimeter has been divided first in two subsystems: the lower part noted (I) and the upper part noted (S). They are separated by a heat radiation shield named H.

For the same reasons given previously the parts (I) and (S) have been subdivided in several elements for which the surfaces exchanging heat transfer by radiation are taken as gray bodies, isothermal and perfectly diffuse (i.e. Lambertian surfaces). These elements are indexed by their surface $S_i$, their total hemispherical emissivity $\varepsilon_i$ and their temperature $T_i$.

In part (I) the electrical device under test (DUT) is at temperature $T_1$ and has an external surface $S_1$ which is assumed to be a sphere of 30 mm in diameter. Its total hemispherical emissivity is $\varepsilon_1$. The DUT support and its current leads inside the measuring cell constitute the thermal conduction resistances named respectively $R_{TH1S}$ and $R_{TH1}$.

The total hemispherical emissivity of the internal surface $S_2$ and external surface $S_3$ of the measuring cell are respectively noted $\varepsilon_2$ and $\varepsilon_3$. Their respective temperatures are $T_2$ and $T_3$. The cell thickness constitutes a thermal conduction resistance named $R_{TH23}$.

The measuring cell support and the current leads constitute the thermal resistances of conduction named respectively $R_{TH3S}$ and $R_{TH3}$. Heat exchange by radiation is considered negligible with these elements since their Biot numbers are lower than 0.3.

The emissivity of the radiation shield H surface ($S_H$) is named $\varepsilon_H$ and its temperature $T_H$. This radiation shield decouples heat radiation between the lower part (I) and the upper part (S) of the calorimeter.

The lower part of the radiation shield E is noted EI, its surface $S_{EI}$ has a total hemispherical emissivity $\varepsilon_{EI}$ and a temperature $T_{EI}$. This part is separated from the upper part named ES by a thermal conduction resistance $R_{THEE}$. As for the lower part, the upper part has a temperature $T_{ES}$ with the surface $S_{ES}$ having a total emissivity $\varepsilon_{ES}$.

The lower part of the vacuum tank noted VI is separated with the upper part VS by a thermal conduction resistance recorded $R_{THVV}$. The surfaces, emissivities and temperatures of VI and VS are named with respective indexes VI and VS. The contacts of the elements ES – VS and EI – VI are modeled by thermal conduction resistances respectively $R_{THVES}$ and $R_{THVEI}$.

It is assumed that the longitudinal thermal gradient in the Styrodur® plates is insignificant, thus we did not introduced a thermal conduction resistance between the part (I) and (S) of the Styrodur insulation

The surface $S_0$ of the liquid nitrogen tank is considered at a constant temperature $T_0 = -196.7$ °C (77 K) and have a total hemispherical emissivity $\varepsilon_0$. The liquid nitrogen tank is hold in place to the vacuum tank flange noted C by a stainless steel tube constituting a thermal conduction resistance $R_{TH0C}$. The current leads are thermalized on the surface $S_0$ and are represented by a thermal conduction resistance $R_{TH0}$

The internal surface of the flange is named $S_C$, its emissivity and temperature, are named respectively $\varepsilon_C$ and $T_C$. A thermal resistance $R_{THCV}$ represents the thermal conduction between the flange C and the vacuum tank.

. The upper external surface $S_S$ of the calorimeter formed by a low emissivity insulating film has a temperature $T_S$. The emissivity of that surface is $\varepsilon_S$. The upper part ($S_S$) of the calorimeter external surface is related to the vacuum tank by a thermal résistance $R_{THSVS}$ due to the presence of Styrodur plates. The lower external surface of the calorimeter noted SI has a total hemispherical emissivity $\varepsilon_I$ and a temperature $T_I$. This part is thermally related to the vacuum tank by four thermal resistances connected in parallel $R_{THSVI}$, $R_{TH4}$, $R_{TH5}$ and $R_{TH6}$

The plate that covers the top of the Styrodur plates has a surface named $S_P$ with a total emissivity $\varepsilon_P$ and a temperature $T_P$. The heat transfer between this plate and the Styrodur upper plate is assumed to be negligible. This plate is in contact with the vacuum tank through the thermal resistance $R_{THPV}$.

The heat transfer between the calorimeter elements in steady state regime is shown on Fig. 3 as an equivalent "electrical network". In this network arteries stand for thermal resistances of different type according to the heat transfer model taken into account between two particular elements.

The properties of the materials involved in our model are temperature-dependent. These properties are mainly the thermal conductivity and specific heat capacity of the calorimeter elements, and the total hemispherical emissivity of their surfaces.

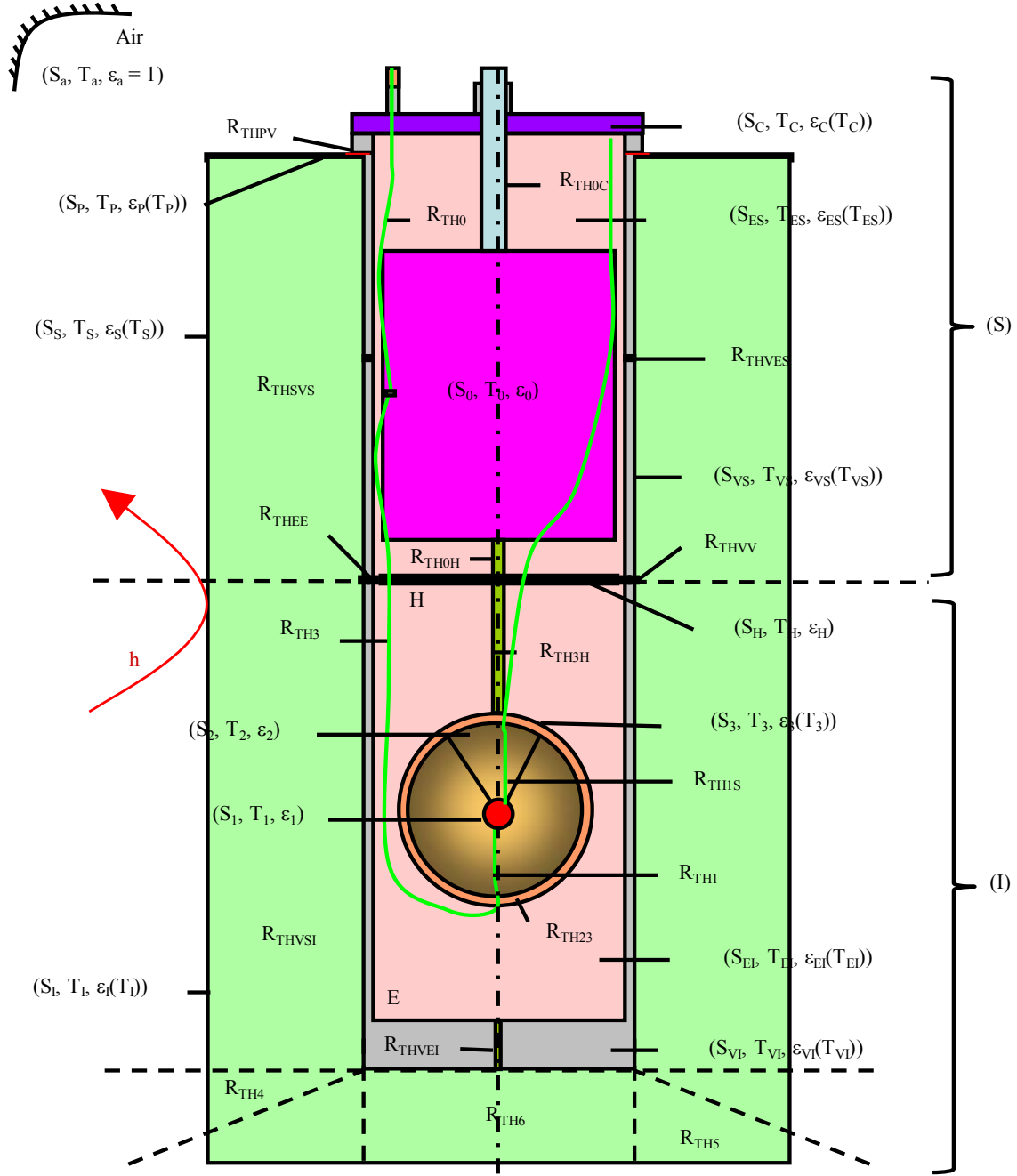

Fig. 2. Index of calorimeter elements taken into account in our modeling of heat transfers. The calorimeter is divided in two systems (I) and (S). These systems interact by thermal conduction resistances $R_{THEE}$ and $R_{THVV}$. Each system is divided himself in subsystems constituted of isothermal elements.

*B. Properties of the materials implemented in the Model*

In this section we describe the thermal properties of materials involved in the calorimeter modeling.

*1) Copper*

The copper used for making our measuring cell ($S_3$) is of OFHC type (Oxygen Free-High Conductivity) [4]. For the thermal conductivity $k_{Cu}$ and specific heat capacity $C_{Cu}$ dependence versus temperature, we used the Simon et al. [5] data available for temperature ranging from 4 K to 300 K. For temperatures above 300 K, we still use the equation from Simon et al. for the thermal conductivity although this latter is not defined in this temperature range. Indeed one can used the Wiedemann-Franz law [6] linking the thermal conductivity and electrical resistivity of copper knowing the temperature resistivity coefficient of the material (e.g. $\alpha = 3.9\ 10^{-3}\ °C^{-1}$ between 0°C and 100 °C).

However this leads to a sharp increase in the thermal conductivity of the copper. Thus, it was interesting to keep going with the equation from Simon et al. [5] for working in the worst case by lowering the thermal conductivity.

*2) Constantan*

In our model of heat transfer, we chose to approximate the thermal conductivity of the constantan by linearizing the Sundqvist data [7]. The thermal conductivity $k_C$ of the constantan is then given by equ. (1).

$$k_C = 19.5 \, [1 + 1.907 \cdot 10^{-3}(T-200)] \quad (1)$$

where T represents the temperature in Kelvin (K).

As shown by Obame *et al.* [2], equ. (1) shows a good agreement with the data of Sundqvist [7]. For electrical resistivity, the measurements performed by Sundqvist show that the average resistivity value of the constantan is 47.8 µΩ.cm at 295 K and is almost independent of temperature.

*3) Gold*

In the calorimeter gold is only deposited as a thin layer on some specific surfaces. It is used for lowering the thermal emissivity and ensuring the protection of surfaces against oxidation. In the case of heat transfer by radiation, we focused only on its thermal emissivity. The Sievers theory [8] enables to calculate the emissivity of pure metallic materials from equations (2) and (3).

$$\varepsilon_N = 2.49 \cdot 10^{-8} \, \rho_{Au} / |R_H|^{1/2} + 3.55 \cdot 10^{-6} \, \rho_{Au} / |R_H|^{1/3} \quad (2)$$

$$\varepsilon_{Au} = 4\varepsilon_N / 3 \quad (3)$$

Fig. 3. Schematic representation of heat transfers between the isothermal elements subdividing the calorimeter: $T_i$ is the temperature of the element i, $\Im_{ij}$, $\Re_{ij}$ and $R_{THij}$ are respectively the global view factor of gray surface ($1/\Im_{ij} = r_i + \Re_{ij} + r_j$, with $r_i = (1-\varepsilon_i)/S_i\varepsilon_i$), the radiation resistance ($\Re_{ij} = 1/F_{ij}S_i$ with $F_{ij}$ the view factor of the element I to element j) and the thermal conduction resistance between element i and j. $Q_1$ and $Q_3$ are the heat powers provided respectively by the DUT when it is switched on and by the measuring cell ($S_3$) regulation.

where $\varepsilon_N$ and $\varepsilon_{Au}$ are respectively the total normal and hemispherical emissivity of the metal, $R_H$ is its Hall constant and $\rho_{Au}$ its electrical resistivity expressed in µΩ.cm.

The electrical resistivity value of gold $\rho_{Au}$ at 293 K is 2.44 µΩ.cm [9] [10]. This electrical resistivity as a function of the temperature can be calculated from equ. (4).

$$\rho_{Au}(T) = \rho_0 \, [1 + \alpha \, (T-T_0)] \quad (4)$$

where, $\rho_0$ is the electrical resistivity at 293 K and $\alpha$ the temperature coefficient in $K^{-1}$. This latter is obtained from Glenn Elert data [11] for temperatures varying from 173 K to 327 K: $\alpha = 3.78 \cdot 10^{-3} \, K^{-1}$.

Equ. (4) introduces an error lower than 1 % in the gold resistivity for temperatures between 200 K and 500 K, thus including our operating range [2]. Then, this equation has been used for the calculation of gold emissivity.

In the model of free electrons used by Sievers [8], the Hall constant $R_H$ is given by equ. (5) [12] [13].

$$R_H = 1/(n \times e) \qquad (5)$$

where n represents the charge carriers density (n = 5.90 $10^{22}$ cm$^{-3}$ for gold) and e the electronic charge (e ≈ -1.6 $10^{-19}$ C).

*4) Stainless steel*

The stainless steel used in our calorimeter is of type 304L. Its thermal conductivity as a function of temperature in the range 4 K to 300 K is given by Touloukian *et al.* [14]. For temperature varying from 300 K to 600 K, the thermal conductivity $k_{St}$ is given by Graves *et al.* [15] formula:

$$k_{St}(T) = 7.9318 + 0.023051\, T + 6.4166\, 10^{-6}\, T^2 \qquad (6)$$

where T represents the temperature in K and $k_{St}$ is expressed in W m$^{-1}$ K$^{-1}$.

The electrical resistivity versus temperature of our steel is given by equ. (4). The constants are determined from Roger data [16] such that $\rho_0$ = 77.41 $10^{-6}$ Ω.cm at $T_0$ = 340 K and α = 9.8906 $10^{-4}$ K$^{-1}$.

In the case of heat radiation, the total hemispherical emissivity of the 304L stainless steel can be calculated from the classical theory of emission proposed by Davidson & Weeks [16] [17]. Thus the emissivity is given by

$$\varepsilon_{St}(T) = a\,(\rho_{St}T)^{1/2} - b\,(\rho_{St}T) + c\,(\rho_{St}T)^{3/22} - d\,(\rho_{St}T)^2 \qquad (7)$$

where a = 0.754, b = 0.632, c = 0.670 and d = 0.607. The quantities $\rho_{St}$ and T are respectively the electrical resistivity of the steel in Ω.cm and the temperature in K. The precision of this expression is 1 % if the product $\rho_{St}T$ is lower than 0.1 Ω.cm.K. which is our case for temperatures varying from 100 K to 500 K.

*5) Glass fiber epoxy composite material*

The thermal conductivity $k_S$ of glass fiber epoxy composite material used in the model is given in reference [2] [18] [19] [20]. It is a temperature dependent function.

*6) Styrodur®*

The thermal conductivity of the styrodur plate is $k_{Styr}$ = 0.034 W.m$^{-1}$.K$^{-1}$. In our model the thermal conductivity of the styrodur is considered as a constant value.

C. *Expressions of the different View factors*

The view factor in the case of diffuse emitting surfaces is defined as the ratio of heat radiated by a surface $S_i$ which reaches the surface $S_j$ to the total heat radiated by $S_i$.

The calorimeter surfaces being considered as perfectly diffuse, we give in this section the different view factors involved in the model.

*1) View factors between $S_0$-$S_{ES}$, $S_0$-$S_H$ and $S_H$-$S_{ES}$*

The view factor $F_{0,H}$ between the surfaces $S_0$ and $S_H$ (i.e. between the liquid nitrogen tank surface and the radiation shield H surface, see Fig. 4) is calculated as the sum of $F_{01,H}$ and $F_{02,H}$. The view factor $F_{01,H}$ is determined from Feingold relation [21] given by equ. (8).

$$F_{01,H} = \tfrac{1}{2}\{X - [X^2 - 4(R_H/R_0)^2]^{1/2}\} \qquad (8)$$

where $R_H = r/a_H$, $R_0 = r_0/a_H$ and $X = 1+(1 + R_H^2 / R_0^2)$.

Let $F_{02,F}$ be the view factor of $S_{02}$ to $S_F$. $F_{02,F}$ is given by Samuel Rea expression [22]. The view factor $F_{02,H}$ is then determined as the half of the quantity 1-$F_{02,F}$.

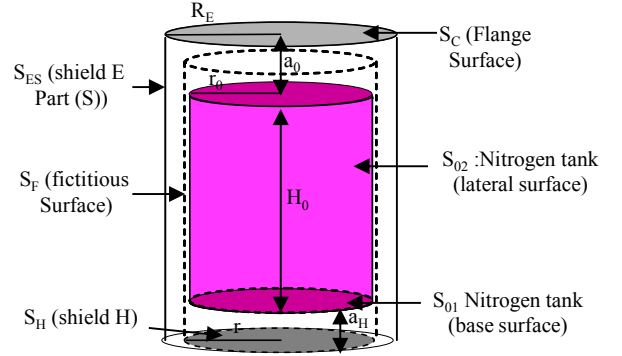

Fig. 4.  Schema of facing surfaces around the liquid nitrogen tank surface to determine the different expressions of view factors.

Therefore $F_{0,H}$ is given by equ. (9).

$$F_{0,H} = \tfrac{1}{2}(1 - F_{02,F}) + F_{01,H} \qquad (9)$$

The calculation of view factor $F_{0,C}$ of $S_0$ to $S_C$ (nitrogen tank surface to vacuum tank flange) is determined in a same way as that of $F_{0,H}$. However one must replace, in this case, r by $R_E$ (vacuum tank radius) and $a_H$ by $a_0$ (see Fig. 4).

The view factor $F_{0,ES}$ of $S_0$ toward $S_{ES}$ is deduced from the previous view factors such that:

$$F_{0,ES} = 1 - (F_{0,H} + F_{0,C}) \qquad (10)$$

The view factor of $S_H$ to $S_C$ is taken as zero in our model since these surfaces are hidden each other by the nitrogen vessel ($S_0$).

Owing to the short distance between $S_{ES}$ and $S_{VS}$ on one hand, and $S_{EI}$ and $S_{VI}$ on the other one, the view factors $F_{ES,VS}$ and $F_{EI,VI}$ are almost equal to unity. The view factors $F_{ES,VI}$ and $F_{EI,VS}$ are considered equal to zero.

*2) View factors between $S_3$-$S_H$, $S_3$-$S_{EI}$ and $S_H$-$S_{EI}$*

The view factor of the surface $S_3$ toward the surface $S_H$ is calculated from [23] [24] and is given by equ. (11).

$$F_{3,H} = \tfrac{1}{2}\{1 - 1/(1 + R^2)^{1/2}\} \qquad (11)$$

where R is equal to the ratio r/*a* in which r represents the radius of the shield H and *a* the distance from the shield H to the center of the sphere $S_3$ (i.e. the measuring cell).

The view factors $F_{3,EI}$ of $S_3$ toward $S_{EI}$, and $F_{H,EI}$ between $S_H$ and $S_{EI}$ are calculated respectively from (12) and (13).

$$F_{3,EI} = 1 - F_{3,H} \qquad (12)$$

$$F_{H,EI} = 1 - F_{3,H} \times S_3 / S_H \qquad (13)$$

The view factor $F_{1,2}$ of the DUT toward the measuring cell is equal to unity. This is also the case for the view factors $F_{SI,a}$, $F_{SS,a}$ and $F_{C,a}$ of the respective surfaces $S_I$, $S_S$ and $S_C$ toward the room environment. This latter is taken as a closed surface ($S_a$, $\varepsilon_a$) around the calorimeter and maintained at constant temperature $T_a$ ($T_a = 300$ K in most cases).

*D. Calorimeter heat transfer modeling*

The analytical equations based on heat balance take into account the three heat transfer modes. In this section, all the equations are carried out in a steady state regime.

In the next sections the term denoted $T_{i,j}$ represents the half sum of the temperatures $T_i$ and $T_j$.

*1) Heat balance for the DUT*

Heat balance applied to the DUT is set in equation (14).

$$\mathfrak{I}_{1,2}\sigma(T_1^4-T_2^4)+(T_1-T_2)(1/R_{TH1}+1/R_{TH1S})-Q_1 = 0 \quad (14)$$

The parameter $\sigma$ represents the Stefan-Boltzmann constant ($\sigma = 5.67\ 10^{-8}$ W m$^{-2}$ K$^{-4}$).

The symbol $\mathfrak{I}_{1,2}$ represents the global radiation factor [2] [25] which takes into account the view factor, the heat exchanging surfaces areas and their thermal emissivities:

$$\mathfrak{I}_{1,2} = (1-\varepsilon_1)/(\varepsilon_1 S_1)+1/(F_{1,2}S_1)+(1-\varepsilon_2)/(\varepsilon_2 S_2) \quad (15)$$

In the model, the DUT surface emissivity $\varepsilon_1$ is assumed to be independent of temperature but its value can vary between 0.1 and 0.9. The internal surface emissivity $\varepsilon_2$ value of the measuring cell is taken equal to that of the Nextel® velvet coating 811-21 black paint (i.e. $\varepsilon_2 = 0.97$).

The factor $\mathfrak{I}_{i,j}$ will conserve the same definition for some following equations by changing only indexes i and j.

The thermal conduction resistances $R_{TH1}$ and $R_{TH1S}$ are expressed respectively by equ. (16) and (17).

$$R_{TH1} = l_1/\{2\times k_C(T_{1,2})\times A_1\} \quad (16)$$

$$R_{TH1S} = l_{1S}/\{2\times k_S(T_{1,2})\times A_{1S}\} + l_{SA}/\{k_S(T_1)\times A_{AS}\} \quad (17)$$

$l_1 = 110$ mm represents the length of current leads inside the measuring cell, $A_1 = 0.785$ mm$^2$ its cross section and $k_C$ the thermal conductivity of the constantan material. $l_{1S} = 77$ mm and $l_{SA} = 3$ mm are respectively the length of the component support with a cross section $A_{1S} = 0.196$ mm$^2$ and the component attachment system with a cross section $A_{AS} = 2$ mm$^2$. The quantity $k_S$ represents the thermal conductivity of glass fiber epoxy composite material. $Q_1$ represents the heat power generated by the DUT when it is switched on in thermal steady state regime.

*2) Heat balance for the Measuring cell*

The thermal balance applied between the internal surface (radius $R_2 = 90$ mm) and the external surface (radius $R_3 = 100$ mm) of the measuring cell is given by equ. (18).

$$Q_1 + (T_2-T_3)/R_{TH23} = 0 \quad (18)$$

$R_{TH23}$ represents the thermal conduction resistance of a copper hollow sphere with an internal radius $R_2$ and an external radius $R_3$.

Equ. (19) below is obtained by applying the heat balance for the measuring cell when radiation transfers of the cell support is not considered[1].

$$Q_1 + Q_3 + \sigma(T_{EI}^4-T_3^4)/\mathcal{R}_{3,EI} + \sigma(T_H^4-T_3^4)/\mathcal{R}_{3,H}$$

$$+(T_0-T_3)(2/R_{TH3}+1/R_{TH3S}) = 0 \quad (19)$$

$Q_3$ represents the heat power generated by the measuring cell temperature controller. $\mathcal{R}_{3,EI}$ and $\mathcal{R}_{3,H}$ are the equivalent radiation resistances of respectively the exchange between measuring cell and radiation shield E and the exchange between measuring cell and radiation shield H. These resistances are deduced from the Kennely transformations. They are functions of view factors, radiated surface areas and surface emissivities.

The resistances of conduction $R_{TH3}$ is defined in the same way as $R_{TH1}$ by replacing the corresponding indexes ($l_3 = 0.63$ m, $A_3 = A_1$). However $R_{TH3S}$ is given by equ. (20) with $l_{3S} = 0.18$ m and $A_{3S} = 6.28\ 10^{-5}$ m$^2$.

$$R_{TH3,S} = l_{3,S}/\{k_S(T_{3,0})\times A_{3,S}\} \quad (20)$$

*3) Heat balance for the Radiation shield E*

  *a) Lower part of E*

The heat balance on the lower part of the radiation shield E is given by equ. (21).

$$\sigma(T_{EI}^4-T_3^4)/\mathcal{R}_{3,EI} + \sigma(T_{EI}^4-T_H^4)/\mathcal{R}_{H,EI} + \mathfrak{I}_{EI,VI}(T_{EI}^4-T_{VI}^4)$$

$$+(T_{EI}-T_{VI})/R_{THE,VI} + (T_{EI}-T_{ES})/R_{THE,E} = 0 \quad (21)$$

The radiation resistance $\mathcal{R}_{H,EI}$ is deduced from the Kennely transformations (simultaneously with the resistances $\mathcal{R}_{3,EI}$ and $\mathcal{R}_{3,H}$). The conduction resistances $R_{THE,VI}$ composing the shield E support is defined by substituting in equ. (20) the indexes 3 and S respectively by E and VI. Its length is $l_{E,VI} = 5\ 10^{-2}$ m and its cross section $A_{E,VI} = 1.96\ 10^{-5}$ m$^2$. $R_{THE,E}$ is the conduction resistance of a hollow cylinder separating the part (S) and (I) of the shield E and having the following characteristics: a length $l_{E,E} = 10$ mm, an internal radius of 129 mm, a thickness of 1 mm and a cross section $A_{E,E} = 2.03\ 10^{-4}$ m$^2$.

---

[1] This is an important hypothesis which leads to great simplifications. This hypothesis is more or less important depending on the support cell material and will only be used for section IV. For section V, transfers by radiation of the cell support (with their corresponding view factors) have been introduced into equ. (19) in order to make a comparison with the experimental results using the most complete possible configuration.

*b) Upper part of E*

The equivalent "electrical circuit" for the heat transfer by radiation on the upper part of the radiation shield E (ES) is shown on Fig. 5 for a thermal steady state regime. The heat transfers by radiation are performed between the four surfaces $S_{ES}$, $S_C$, $S_H$ and $S_0$ (see Fig. 2).

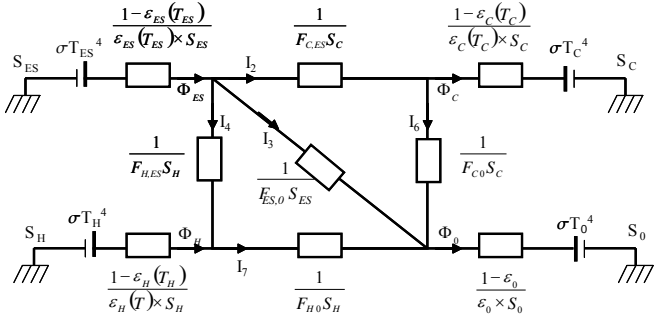

Fig. 5. Electrical analogical schema of heat radiation between the gray surfaces $S_{ES}$, $S_C$, $S_0$ and $S_H$.

The resulting system of equations has been solved to determine the expressions of the following outlet heat radiation fluxes: $\Phi_{ES}$ from $S_{ES}$, $\Phi_C$ from $S_C$, $\Phi_H$ from $S_H$ and $\Phi_0$ from $S_0$. The calculated fluxes are functions of the emitting surfaces properties (i.e. surface areas, surface emissivities, view factors and temperatures).

On the upper part of the emitting surface $S_{ES}$, the heat balance leads to equ. (22).

$$\Phi_{ES} + \Im_{ES,VS}\,\sigma(T_{ES}^4 - T_{VS}^4) + (T_{ES} - T_{VS})/R_{THES,VS}$$

$$+ (T_{ES} - T_{EI})/R_{THE,E} = 0 \qquad (22)$$

$R_{THES,VS}$ is the equivalent conduction resistance due to the glass fiber epoxy composite rods maintaining the shield E upright. This resistance is calculated from the equ. (20) with $l_{ES,VS} = 10^{-2}$ m and $A_{ES,VS} = 7.85\,10^{-5}$ m$^2$.

*4) Heat balance for the Radiation shield H*

The heat balance applied to the radiation shield H leads to equ. (23).

$$\Phi_H + \sigma(T_H^4 - T_3^4)/R_{3,H} + \sigma(T_H^4 - T_{EI}^4)/R_{H,EI}$$

$$+ (T_H - T_0)/R_{TH0,H} + (T_H - T_3)/R_{TH3,H} = 0 \qquad (23)$$

The conduction resistances $R_{TH0,H}$ and $R_{TH3,H}$ are again determined from equ. (20) using their respective lengths $l_{H,0} = 14\,10^{-2}$ m and $l_{3,H} = 4\,10^{-2}$ m, and cross sections such that $A_{H,0} = A_{3,H} = A_{3,S}$.

*5) Heat balance for the Vacuum tank*

*a) Upper part of the vacuum tank (VS)*

The heat balance equation is applied to the upper part of the vacuum tank and leads to equ. (24).

$$\Im_{ES,VS}\,\sigma(T_{ES}^4 - T_{VS}^4) + (T_{ES} - T_{VS})/R_{THES,VS}$$

$$+ (T_{VI} - T_{VS})/R_{THV,V} + (T_S - T_{VS})/R_{THS,VS}$$

$$+ (T_C - T_{VS})/R_{THVS,C} + (T_P - T_{VS})/R_{THV,P} = 0 \qquad (24)$$

The conduction resistance $R_{THV,V}$ and $R_{THVS,C}$ are calculated using equ. (20) with the respective indexes: $l_{V,V} = l_{VS,C} = 10^{-2}$ m, $A_{V,V} = A_{VS,C} = 1.746\,10^{-3}$ m$^2$. The resistance $R_{THS,VS}$ is equivalent to the thermal resistance of a cylindrical tube having an equivalent external radius of $R_{eq}$ (the calorimeter external radius) and the vacuum tank external radius $R_V = 0.14$ m. The length of the tube $L_S = 0.465$ m is that of the upper part of the vacuum tank.

The contact between the vacuum tank and the Styrodur stainless steel top plate is ensured by a PTFE seal and forms a conduction resistance $R_{THV,P}$ determined also from the equ. (20) with $l_{V,P} = 2\,10^{-3}$ m, $A_{Tef} = 1.1\,10^{-2}$ m$^2$ and a thermal conductivity $k_{Tef} = 0.21$ W m$^{-1}$ K$^{-1}$ considered as a constant value.

*b) Lower part of the vacuum tank (VI)*

The heat balance applied to the lower part of the vacuum tank provides equ. (25).

$$\Im_{EI,VI}\,\sigma(T_{EI}^4 - T_{VI}^4) + (T_{EI} - T_{VI})/R_{THE,VI} + (T_{VS} - T_{VI})/R_{THV,V}$$

$$+ (T_I - T_{VI})(1/R_{THV,SI} + 1/R_{TH4} + 1/R_{TH5} + 1/R_{TH6}) = 0 \quad (25)$$

The thermal resistance $R_{THV,SI}$ has the same geometry properties as the resistance $R_{THS,VS}$ excepted that its tube length is $L_I = 0.495$ m.

The thermal resistances $R_{TH4}$, $R_{TH5}$ and $R_{TH6}$ are due to the Styrodur bottom plate (see Fig. 2). The resistance $R_{TH6}$ corresponds to the thermal conduction resistance of a full cylinder and is given by equ. (26).

$$R_{TH6} = e/(k_{Styr} \times \pi \times R_V^2) \qquad (26)$$

where $e = 0.1$ m represents the thickness of the Styrodur plate and $R_V$ is the vacuum tank external radius.

The calculation of the two resistances $R_{TH4}$, $R_{TH5}$ has been done by analytical integration of a triangular section of revolution.

*6) Heat balance on the flange of the Vacuum tank (C)*

The heat exchange on the flange of the vacuum tank is given by equ. (27).

$$\Phi_C + \Im_{C,a}\,\sigma(T_a^4 - T_C^4) + (T_0 - T_C)(1/R_{TH0,C} + 2/R_{TH0})$$

$$+ (T_{VS} - T_C)/R_{THC,V} + h_C\,S_C(T_a - T_C) = 0 \qquad (27)$$

The room environment emissivity is assumed to be equal to 1 (i.e. it behaves as a black body). The equivalent thermal resistance $R_{TH0,C}$ is given by equ. (28).

$$R_{TH0,C} = 1 / k_{St}(T_{0,C})\{l_{0,C1}/A_{0,C1} + l_{0,C2}/A_{0,C2}\} \quad (28)$$

where $k_{St}$ is the thermal conductivity of stainless steel. $A_{0,C1}$ and $A_{0,C2}$ are the cross sections of the two resistances connected in series which represent the support of the nitrogen tank. The length of these resistances are $l_{0,C1} = 0.2$ m and $l_{0,C2} = 0.08$ m, and their cross sections are $A_{0,C1} = 7.54 \cdot 10^{-5}$ m$^2$, $A_{0,C2} = 1.2 \cdot 10^{-4}$ m$^2$.

The expression of the resistance $R_{TH0}$ is given by equ. (29).

$$R_{TH0} = l_0 /\{k_{Cu}(T_{0,C}) \times A_1\} + l_{01} /\{k_{Cu}(T_C) \times A_T\}$$

$$+ l_{02} /\{k_{St}(T_C) \times A_T\} \quad (29)$$

$l_0$ represents the length of current leads from the thermalization point located on the liquid nitrogen tank to the tulip placed on the vacuum tank flange [2] [3]. $A_1$ is its cross section. $l_{01} = 0.05$ m represents the length of the copper part of the tulip and $l_{02} = 0.08$ m is that of the stainless steel part and their cross section is $A_T = 1.65 \cdot 10^{-3}$ m$^2$.

$R_{THC,V}$ is the resistance due to the flange thickness and is expressed by replacing in equ. (20) the indexes 3 and S respectively by C and V on the one hand and $k_S(T_{3,S})$ by $k_{St}(T_{C,V})$ on the other one.

$h_C$ is the natural convection coefficient of air. Several mathematical relations of this coefficient exist in literature, but in our model we used the Brau *et al.* [27] simple relation given by equ. (30).

$$h_C = 1.78 \times \Delta T^{0.25} \quad (30)$$

where $\Delta T = |T_C - T_a|$ represents the difference of temperature between the air at room temperature and the calorimeter wall which is in the present case the flange of the vacuum tank.

*7) Heat balance for the Styrodur top cover*

By applying the heat balance on the Styrodur top cover with surface $S_P$ (see Fig. 2) we obtained equ. (31).

$$\Im_{P,a} \sigma(T_\alpha^4 - T_P^4) + (T_{VS} - T_P) / R_{THP,V} + h_P S_P (T_a - T_P) = 0 \quad (31)$$

The coefficient $h_P$ is determined from equ. (30).

*8) Heat balance for the insulating film covering the calorimeter enclosure*

The heat balance applied on the upper (SS) and the lower (SI) parts of the low thermal emissivity insulation film covering the calorimeter gives respectively the relations (32) and (33).

$$\Im_{S,a} \sigma(T_\alpha^4 - T_S^4) + (T_{VS} - T_S)/R_{THS,VS} + h_S S_S (T_a - T_S) = 0 \quad (32)$$

$$\Im_{I,a} \sigma(T_\alpha^4 - T_I^4) + (T_{VI} - T_I)(1/R_{THV,SI} + 1/R_{TH4} + 1/R_{TH5}$$

$$+ 1/R_{TH6}) + h_I S_I (T_a - T_I) = 0 \quad (33)$$

The coefficients $h_S$ and $h_I$ are determined from the equ. (30).

*E. Order of magnitude of conduction resistances*

Orders of magnitude of conduction resistances used in the model are given in TABLE I. The resistances being dependent on the temperature; their calculations are carried out at a constant temperature of 300 K [2].

TABLE I. ORDERS OF MAGNITUDE OF THE DIFFERENT THERMAL CONDUCTION RESISTANCES. AT 300 K

| Conduction resistance | Order of magnitude [W/K] |
|---|---|
| $R_{TH1}$ | $1.80 \cdot 10^2$ |
| $R_{TH1S}$ | $1.17 \cdot 10^4$ |
| $R_{TH23}$ | $4.54 \cdot 10^{-4}$ |
| $R_{TH3}$ | $1.03 \cdot 10^3$ |
| $R_{TH3S}$ | $3.6 \cdot 10^3$ |
| $R_{THEI,VI}$ | $3.2 \cdot 10^3$ |
| $R_{THE,E}$ | $1.3 \cdot 10^{-1}$ |
| $R_{THES,VS}$ | $0.53 \cdot 10^2$ |
| $R_{TH0,H}$ | $7.96 \cdot 10^2$ |
| $R_{TH3,H}$ | $2.79 \cdot 10^3$ |
| $R_{THV,V}$ | $3.8 \cdot 10^{-1}$ |
| $R_{THVS,C}$ | $3.8 \cdot 10^{-1}$ |
| $R_{THS,VS}$ | $2.44$ |
| $R_{THP,V}$ | $8.6 \cdot 10^{-1}$ |
| $R_{THV,SI}$ | $1.19 \cdot 10^1$ |
| $R_{TH4}$ | $1.63 \cdot 10^2$ |
| $R_{TH5}$ | $3.07 \cdot 10^1$ |
| $R_{TH6}$ | $4.77 \cdot 10^1$ |
| $R_{TH0,C}$ | $2.21 \cdot 10^2$ |
| $R_{TH0}$ | $6.26 \cdot 10^2$ |
| $R_{THC,V}$ | $2.7 \cdot 10^{-2}$ |

## IV. ANALYSIS OF THE RESULTS FROM THE MODELING

In our simulations the support of the measuring cell (thermal resistance $R_{TH3,S}$) is made of glass fiber epoxy composite [2] excepted in the indicated cases. The temperature $T_3$ of the measuring cell surface $S_3$ will be called the measuring cell temperature. The temperature of the calorimeter external environment is called here ambient room temperature (or simply ambient temperature).

*A. Influence of the ambient room temperature*

The ambient room temperature influence on the calorimeter is highlighted by a calculation aiming to observe the inlet heat fluxes entering the calorimeter and the measuring cell versus the ambient room temperature.

The simulations are achieved here without heat generation in the calorimeter enclosure ($Q_1 = 0$, $Q_3 = 0$). The emissivity of the radiation shield H is fixed to the value 0.1 and the emissivity of the DUT is fixed to the unit value.

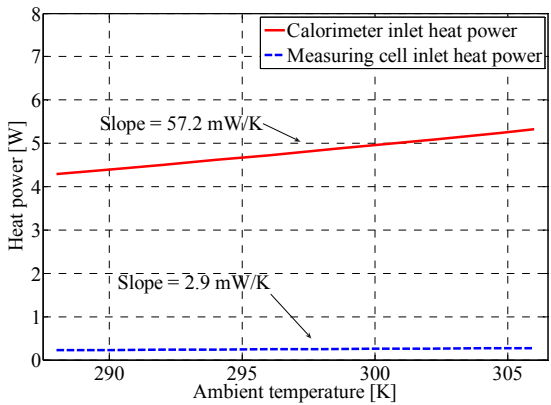

Fig. 6. Evolution of the inlet heat flux into the calorimeter and the measuring cell ($S_3$) versus the ambient temperature ($Q_1 = 0$, $Q_3 = 0$).

On Fig. 6 it is seen that the calorimeter and the measuring cell inlet heat power are evaluated respectively at 4.96 W and 0.26 W when ambient temperature is fixed at 300 K. The calorimeter inlet flux is logically higher than that entering in the measuring cell, the most of heat power flowing to the cooling point (i.e. to the liquid nitrogen tank).

One can also observe that both the calorimeter and the measuring cell inlet heating powers increase when ambient temperature increases and have respective slopes of 57.2 mW $K^{-1}$ and 2.9 mW $K^{-1}$. Measurement being performed on the measuring cell, this shows that the precision of the calorimeter may be affected by small variations of the ambient room temperature.

Fig. 7 shows the evolution of the measuring cell temperature versus the ambient temperature for three different values of the heating power $Q_3$ provided by the temperature controller to maintain the measuring cell respectively at 475 K, 300 K and 225 K.

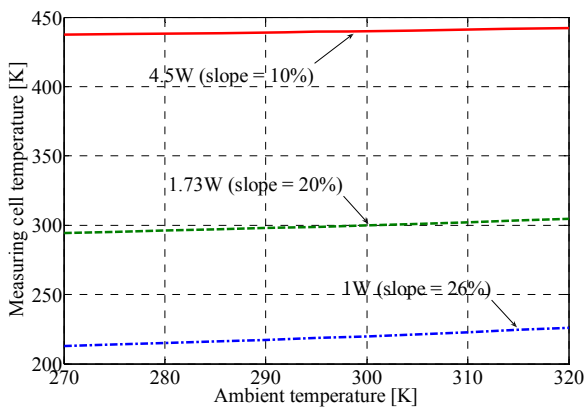

Fig. 7. Temperature variations of the external cell surface $S_3$ versus the ambient room temperature for different heat power $Q_3$ provided by the regulation when the DUT is switched off ($Q_1 = 0$).

In all cases, the measuring cell temperature increases by increasing the ambient temperature. However the increase observed becomes smaller when the controller heating power $Q_3$ becomes larger. Therefore the disturbance brought by the ambient temperature is lowered when the measuring cell operating temperature increases.

*B. Thermal cartography of the calorimeter*

On Fig. 8, the temperature of different elements is shown for a variation of the controller heating power supply $Q_3$ provided to the measuring cell $S_3$. The radiation shield (H) emissivity is always fixed to the value $\varepsilon_H = 0.1$ (i.e. polished stainless steel). The ambient temperature is fixed to 300 K for all the calculations.

In the absence of heat generation in the calorimeter the equilibrium temperature of the measuring cell reaches the value of 225 K when the liquid nitrogen tank is filled ($T_0 = 77$ K).

It is observed that the measuring cell temperature evolves by increasing the controller heating power $Q_3$. Furthermore the temperature of the upper and the lower part of the vacuum tank ($S_{VS}$ and $S_{VI}$) as well as inside the Styrodur ($S_S$ and $S_I$) remain almost constant in all the range of the controller heating power provided.

One can also observe that the temperature of the radiation shields E and H evolve significantly with the heating power supplied by the regulation $Q_3$.

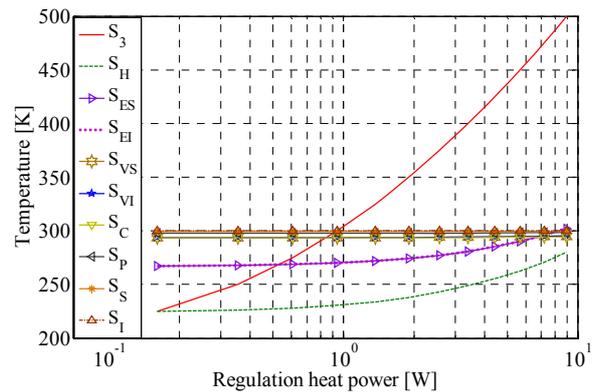

Fig. 8. Temperature variations of the calorimeter's main elements versus the heat power regulating the measuring cell $Q_3$. The DUT is switched off ($Q_1 = 0$).

*C. Influence of the radiation shield H*

To examine the influence of the radiation shield H on the temperature of the measuring cell ($S_3$) a simulation of cell heating by the regulation with and without the shield H is performed. The controller heating power simulated $Q_3$ varies from 1 W to 10 W.

It is observed on Fig. 9 that the temperature increases as it would be expected by increasing the heat power $Q_3$. However, this increase is substantial in the presence of the radiation shield H. This latter favors therefore the increase of the measuring cell temperature and thus increase the heat radiation of the cell by hiding the liquid nitrogen surface ($S_0$) influence.

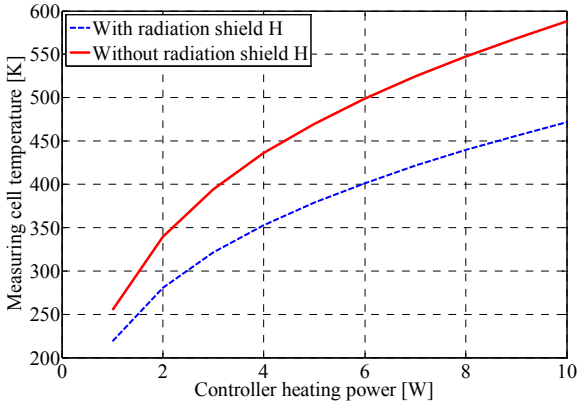

Fig. 9. Temperature evolution of the measuring cell surface $S_3$ versus the heat power $Q_3$ in the presence and in the absence of the heat radiation shield H ($Q_1 = 0$).

The influence of the emissivity $\varepsilon_H$ of the radiation shield H on particular elements is shown on Fig. 10.

It can be observed on Fig. 10 that the temperatures of the measuring cell, of E and H decrease when the emissivity of H increases. However the decrease of the vacuum tank temperature remains weaker than that of other elements. This decrease of temperatures is due to the more intense cooling of H when $\varepsilon_H$ increases.

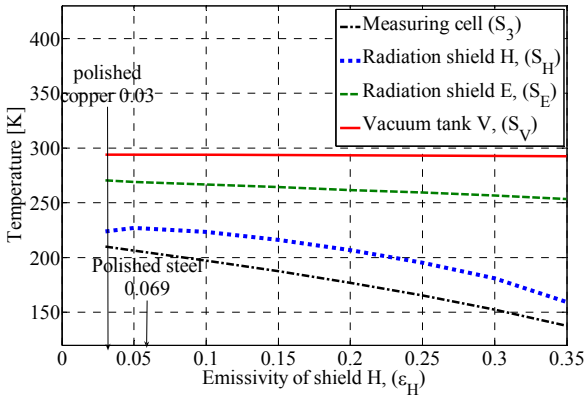

Fig. 10. Influence of the heat radiation shield H emissivity on the temperatures of elements $S_3$, $S_H$, $S_E$ and $S_V$ ($Q_1 = Q_3 = 0$). The ambient room temperature is fixed to 300 K.

Accordingly, the control of the radiation shield H emissivity gives us a possibility to control the minimum achievable temperature of the measuring cell.

### D. Influence of current leads material on the heat radiated by the DUT

The simulations are carried out for the two types of materials used in the current leads: copper and constantan. The heat power $Q_1$ generated by the DUT is fixed at 100 mW. The calculations here are intended to compare the ratio of the radiated heat flux $\Phi_r$ with the total heat flux $\Phi_t$ generated by the DUT (i.e. $\Phi_r/\Phi_t$). The diameter of the current leads is fixed to 1 mm.

For the copper, Fig. 11 shows that the ratio evolves by increasing the measuring cell temperature. It remains lower than 99.5 % up to 380 K. Above this temperature the ratio increases and reaches a value of 99.7 % at 440 K.

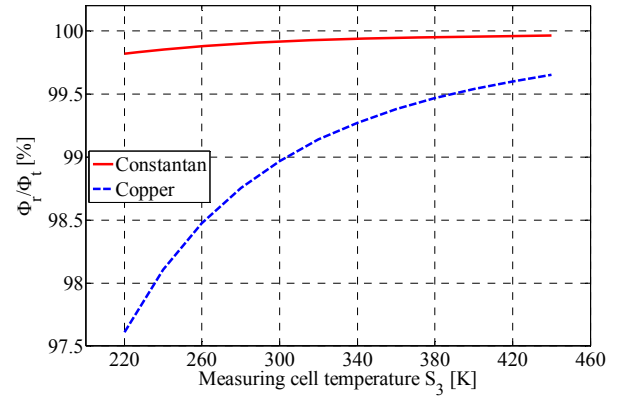

Fig. 11. Evolution of the ratio of the radiated heat flux and the total heat power generated by the DUT versus the temperature of the measuring cell when the current leads are either in copper or in constantan. The total flux dissipated by the DUT is fixed at $Q_1 = 100$ mW. For each cell temperature the regulation provides a $Q_3$ determined by our model.

However for the constantan, the Fig 11 shows a relatively low increase compared to that of the copper. The ratio given by constantan remains above that of copper, it reaches the value of 99.8 % at 220 K and 99.98 % at 440 K.

Therefore the constantan is of great interest since it allows a better heat transfer by radiation compared with copper and it constitutes a good compromise to limit thermal conduction leakage from the current leads. This constitutes one of the criteria for the choice of the DUT material wires knowing that the ratio could be much higher considering the thermal contact resistances between elements.

In all the simulations conducted thereafter, the DUT current leads are then considered in constantan.

### E. Influence of emissivity properties

#### 1) Internal surface of the measuring cell ($S_2$)

To observe the influence of the measuring cell internal surface emissivity $\varepsilon_2$ a simulation of the ratio $\Phi_r/\Phi_t$ is carried out. The calculation of the ratio $\Phi_r/\Phi_t$ versus the cell temperature is performed for different values of $\varepsilon_2$. The total heat flux $Q_1$ dissipated by the DUT is 100 mW, its emissivity is fixed to the unit value and the emissivity of the radiation shield H is fixed to 0.1.

The Fig. 12 shows that for the values of emissivity considered, the ratio increases $\Phi_r/\Phi_t$ with the increase of the cell temperature. In the worst case ($\varepsilon_2 = 0.7$) the ratio is equal at 99.75 %. The heat transfers by radiation are then predominant inside the measuring cell. Experimentally high thermal emissivity, better than 0.97, has been achieved practically by using the Nextel® velvet coating 811-21 [2].

Therefore our measuring cell constitutes a good approximation of a black body.

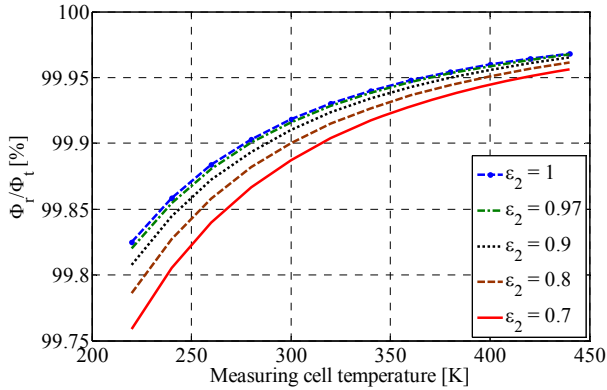

Fig. 12. Evolution of the ratio of the radiated heat flux and the total heat power generated by the DUT versus the temperature of the measuring cell when emissivity of the measuring cell internal surface varies. The total flux dissipated by the DUT is fixed to $Q_1 = 100$ mW. For each cell temperature the regulation provides a $Q_3$ determined by our model.

*2) External surface of the DUT*

The emissivity of the internal surface $S_2$ is fixed now at the value of 0.97 and the total heat dissipated by the DUT remains fixed at 100 mW. Then the ratio $\Phi_r/\Phi_t$ is calculated versus the measuring cell temperature for different values of the DUT surface emissivity.

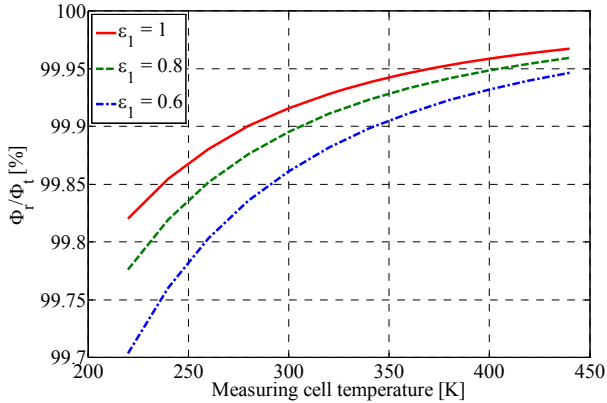

Fig. 13. Evolution of the ratio of the radiated heat flux and the total heat power generated by the DUT versus the temperature of the measuring cell when emissivity of the DUT external surface varies. The total flux dissipated by the DUT is fixed to $Q_1 = 100$ mW. For each cell temperature the regulation provides a $Q_3$ determined by our model.

The Fig. 13 shows that the ratio $\Phi_r/\Phi_t$ increases when increasing the measuring cell temperature. This ratio is improved with the increase of the thermal emissivity of the DUT. One can also observe that the ratio is even higher than 99.7 % when the emissivity is as low as 0.6. This is interesting because DUT have in many cases surfaces cover by polymer or plastic materials. Heat loss measurement with DUT having high emissivity is then suitable when using heat radiation exchange. Another interest is the fact that high emissivity enables to limit the strong rising of DUT surface temperature.

*F. Thermal leakages inside the measuring cell ($S_3$)*

Fig. 14 shows the variations of the heat power dissipated from the measuring cell by conduction through current leads and through the measuring cell support when heat power $Q_3$ is supplied by the regulation to maintain the measuring cell at a fixed temperature. In our calculation $\varepsilon_H$ is fixed to 0.1, the ambient temperature is 300 K and the heating power provided to the measuring cell varies from 0.91 W to 4.47 W. The current wires due to temperature sensors are not taken into account in this simulation.

One can see on Fig. 14 that thermal dissipation through the cell support and current leads is slightly higher than the heating power provided by regulation below 250 K. This is explained by the fact that the measuring cell is also heated by the external environment of the calorimeter. Above 250 K, $Q_3$ starts to be predominant but remains close to the leakages through the wires and cell support up to 300 K. Therefore below 300 K the heat dissipation by conduction is predominant compared to radiation exchange on the surface $S_3$.

However, above 300 K the heating power provided by the regulation becomes increasingly significant compared to the heat leakages through wires supply and cell support. However it remains lower than heat evacuated by conduction through the measuring cell support and current leads in the calorimeter operating range (i.e. from 223 K to 423 K) [2] [3].

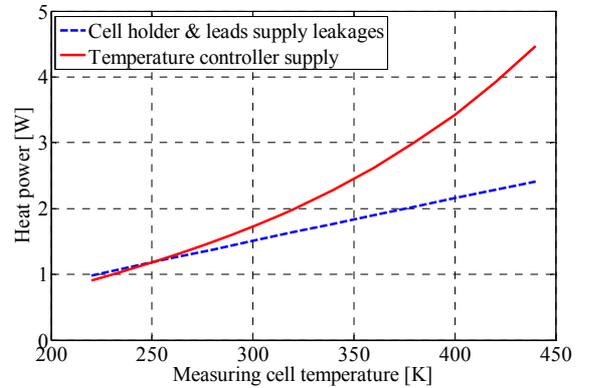

Fig. 14. Evolution of the heat power $Q_3$ provided to the measuring cell to maintain it at temperature $T_3$ and the heat power dissipated by this latter. The total heat power emitted by the DUT is fixed to $Q_1 = 0$.

*G. Thermal dynamic of the calorimeter*

To appreciate the thermal dynamic response of the calorimeter during the cooling (i.e. when the liquid nitrogen tank is filling), we added in each balance equation a term related to heat diffusion in the corresponding element i, of the form $V_i \rho_i C_{Pi} \dfrac{dT_i}{dt}$ where $V_i$, $\rho_i$ and $C_{Pi}$ are respectively the volume, the density and the heat capacity of the element

i. Thus we obtained a Newtonian cooling type differential system of relations.

The Newtonian cooling equation being not relevant with Styrodur (polystyrene) insulation (Biot number Bi = 12.6 >> 1), then the thermal regime of Styrodur is considered as a steady state regime. Fig. 15 shows the temperature evolution of different elements in the calorimeter when the measuring cell support is made of glass fiber epoxy composite material.

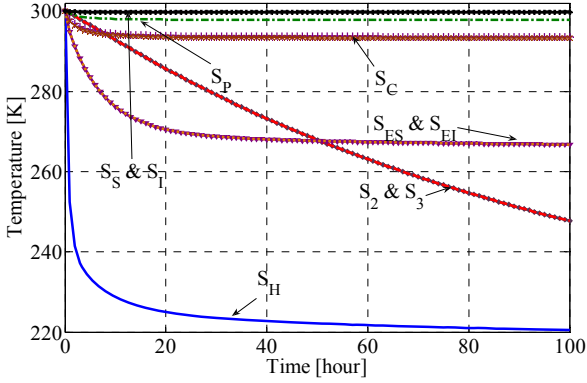

Fig. 15. Time evolution of the temperature of the calorimeter's main elements when no heat generation is considered ($Q_1 = Q_3 = 0$). The support of the measuring cell is made of glass fiber composite epoxy.

It is also observed on Fig. 15 that the stationary regime is reached after long cooling time. The thermal dynamic of the surface shield $S_H$ is logically faster owing to its mass and location in the calorimeter. However the measuring cell dynamic is much slower.

The thermal dynamic of the measuring cell can be improved by using the cell support with better thermal conductivity as can be seen on Fig 16. However this advantage is offset by the increase of experimental measurement inaccuracies.

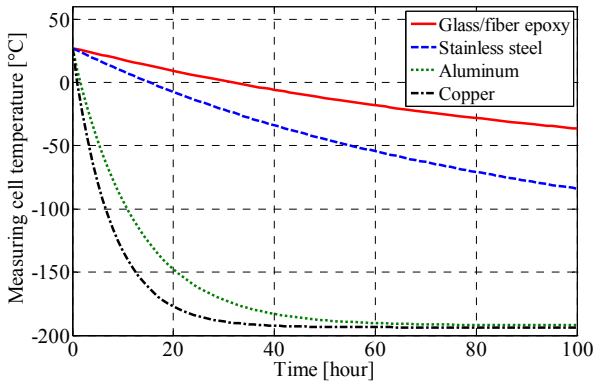

Fig. 16. Time evolution of the measuring cell temperature in the absence of heat generation ($Q_1 = Q_3 = 0$) for different materials constituting the cell support.

## H. Evaluation of the DUT surface temperature

Calculations of the maximal temperature difference $\Delta T_{max}$ have been performed to determine the order of magnitude of the DUT surface temperature.

According to the measurement principle of our calorimeter [2] [3], heating power dissipated by a DUT ($Q_1$) is measurable at a given temperature $T_3$ imposed to the cell only if the controller heating power provided by the regulation $Q_3$ is higher than $Q_1$. The temperature at which we can have $Q_3 = Q_1$ determines then the lowest calorimeter operating temperature for measuring the heat quantity $Q_1$ dissipated by the DUT.

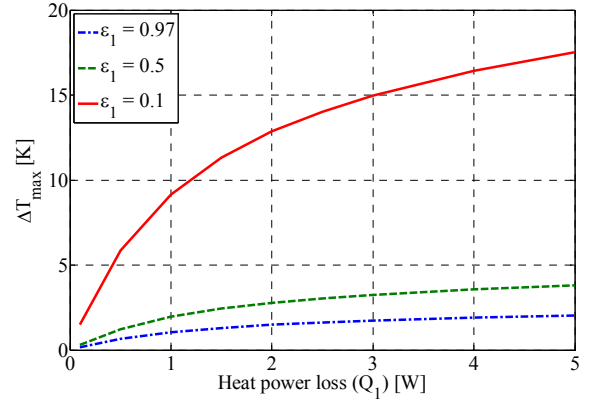

Fig. 17. Maximum temperature difference $\Delta T_{max} = T_1 - T_3$ between the DUT and the measuring cell versus the DUT dissipated heat power loss $Q_1$ for different surface emissivities of the DUT. The support of the measuring cell is made of glass fiber composite epoxy.

Fig. 17 shows the maximum temperature difference $\Delta T_{max}$ between the measuring cell and the DUT versus heat power $Q_1$ for different surface emissivity $\varepsilon_1$ of the DUT. In this simulation the DUT surface is an isotherm. As expected, $\Delta T_{max}$ increases with the increase of both the heat power loss dissipated by the DUT ($Q_1$) and the component surface emissivity. For $\varepsilon_1$ equal to 0.97 and 0.5, the $\Delta T_{max}$ remains respectively lower than 2.5 K and 4 K. However for DUT with low emissivity such as 0.1 (i.e. devices with metallic polished surface), the calculations show a large temperature difference $\Delta T_{max}$ of about 17 K for $Q_1 = 5$ W. This may affect the component temperature operating condition. The solution is to increase the emissivity $\varepsilon_1$ by using for example the Nextel® velvet coating 811-21 black paint ($\varepsilon = 0.97$) to weaken this $\Delta T_{max}$.

## I. Sensibility and performance provided by our modeling

Heat modeling allows to determine the temperature variation $\Delta T_3$ caused by a heat power generated by a DUT ($Q_1$) on the measuring cell maintained at a given temperature $T_3$.

Fig. 18 presents the evolution of $\Delta T_3$ versus the temperature of the measuring cell when $Q_1$ have the following values: 0.1 mW, 1 mW and 10 mW.

In all cases $\Delta T_3$ decreases by increasing the temperature of the measuring cell. However, this variation increases when the heat power generated by the DUT increases.

The TABLE II gives the values taken by $\Delta T_3$ at 200 K and 500 K which are respectively the lowest and the highest operating temperature of our calorimeter.

Thus, for $Q_1 = 0.1$ mW, $\Delta T_3$ at 200 K is equal to 20 mK. The corresponding theoretical variation at 500 K is 1.3 mK.

The measurement of a heat power loss of 0.1 mK for a temperature of 500 K imposed to the measuring cell requires therefore instruments with high resolution of about 1 mK. This can be achieved by using devices such as a PTC 10® of Stanford Research System which shows a resolution corresponding to that value [26] and sensors such as the Pt100.

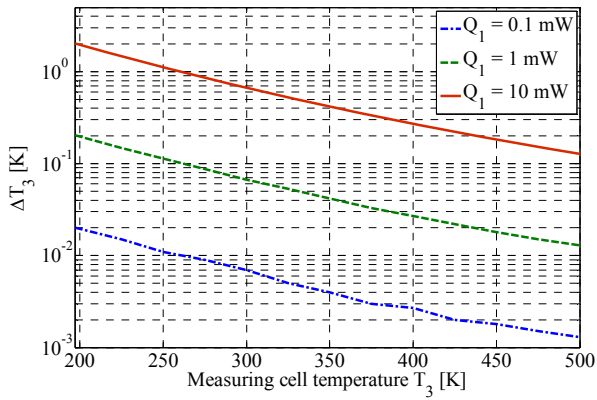

Fig. 18. Evolution of the measuring cell temperature difference $\Delta T_3$ when the DUT changes from inactive to active status depending on the initial temperature of the cell maintained by the regulation. The cell support is made of glass fiber composite epoxy. $Q_1$ takes the different following values: 0.1 mW, 1 mW and 10 mW.

TABLE II. TEMPERATURE DIFFERENCE ARISING ON THE CELL BY HEAT POWER GENERATED BY A DUT FOR TWO INITIAL TEMPERATURES

| Heat power loss $Q_1$ [mW] | $\Delta T_3$ for initial cell temperature of 200 K | $\Delta T_3$ for initial cell temperature of 500 K |
|---|---|---|
| 0.1 | 0.02 | 0.0013 |
| 1 | 0.204 | 0.013 |
| 10 | 2.019 | 0.128 |

V. BRIEF COMPARISON BETWEEN SOME THEORETICAL AND EXPERIMENTAL RESULTS

In this section, we will compare the measuring cell theoretical thermal dynamic response and the calculated heat power loss to that achieved experimentally. These results show preliminary tests giving a first glimpse of the general behavior of our calorimeter. This first analysis is also designed to check whether the original objectives can be achieved with architecture and selected materials.

More results of heat losses measured on different kinds of electrical devices can be found in [28].

*A. Thermal dynamic response of the measuring cell*

The theoretical thermal dynamic response of the measuring cell ($S_3$) when its support is in glass fiber epoxy composite material has been compared to the one obtained experimentally on Fig. 19. It is observed that the theoretical dynamic is slower than that given by the experiment for a cooling time below 55 h. For cooling time above 55 h, one can see the reverse effect. The difference between the two curves reveals the limits of our modeling. Indeed we have not taken into account the thermal resistance introduced by the contacts between the cell and its support and between the support of the cell and the nitrogen tank. Furthermore, the influence of the support capacity has also been not taken into account.

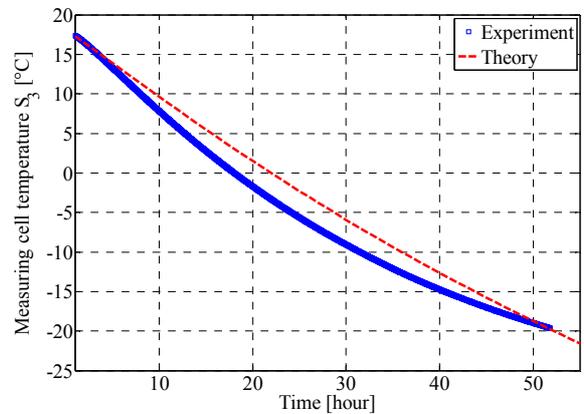

Fig. 19. Time evolution of the measuring cell temperature resulting from the filling of the liquid nitrogen tank. The temperature controller is switched off and there is no DUT in the cell. The temperature has been measured by Pt100 sensors [1]. The cell support is made of glass fiber composite epoxy. The vacuum tank is exhausted at $5 \cdot 10^{-6}$ mbar. This evolution is compared to that obtained theoretically without taking into account the thermal contact resistances between the cell and its support and between the support of the cell and the nitrogen tank.

However, the theoretical curve remains a good approximation for cooling time below 55 h since the difference with experience is lower than 5 K.

*B. Heat power measurement range*

We recall that a measure can be achieved only if the heat power loss dissipated by a DUT ($Q_1$) is lower than the heating power $Q_3$ provided by the regulation maintaining the cell at a fixed temperature $T_3$. Then the heat power $Q_3$ can be considered as the upper limit of measurable heat power loss at a corresponding temperature set point.

Fig. 20 emphasizes the comparison between theoretical and experimental data of the temperature controller heating power $Q_3$. The cell support is in glass fiber epoxy composite material and the calorimeter external temperature is 18 °C.

Heat transfers by radiation of the measuring cell support have been taken into account for simulating the behavior of the calorimeter as it is explained with the heat balance equ. (19).

It is observed that the regulation heating power increases as it would be expected when the temperature increases in both cases. However the experimental curve remains above the theoretical one and the shift between them increases by increasing the temperature set on the cell.

Thus, for a temperature of 80 °C imposed to the cell, the maximal heat power measurable is 4 W in our theoretical approach and 6 W in practice. For a temperature of 100 °C that heat power is 5 W in our theoretical approach and 8 W in practice.

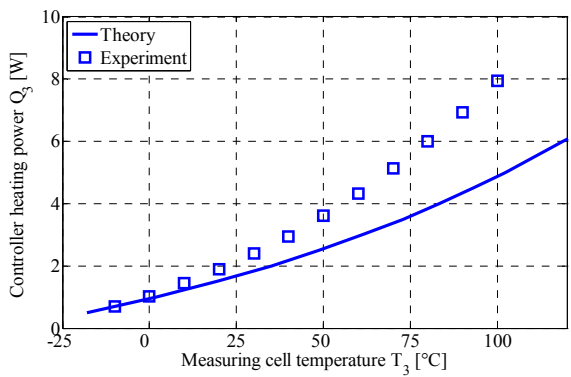

Fig. 20. Comparison between experimental and theoretical heat power provided by the regulation to maintain the measuring cell at fixed temperature. The component is switched off and the cell support material is of glass fiber composite epoxy.

The smallest experimental temperature on Fig. 20 is limited because at temperature lower than –20 °C the regulation heating power is weak for a cell support in glass fiber epoxy composite material thus temperatures below this value require a very long cooling time. The solution adopted for low cell temperature (i.e. for covering the calorimeter operating range) has been the utilization of cell support materials with better thermal conductivities such as stainless steel and aluminum [2].

At temperatures above the room temperature, a possible explanation of the difference observed between our theoretical approach and our experiment is the possibility of more intense gradient of temperatures and thus our subdivision in isothermal elements is not enough. Also thermal dilatation could modify significantly contact thermal resistances of conduction. This argument is supported by the fact that the difference on Fig. 20 between our theoretical approach and our experience is almost linear.

We have already remarked that simplifications have been made in our modeling such that some thermal conduction resistances, which were not *a priori* defined, were not taken into account, e.g. the thermal conduction resistance of current leads for the Pt100 sensors.

However, the agreement below room temperature between our theoretical approach and the experiment is an important result because the possibility to perform measurements in this low temperature region is an improvement of our calorimeter.

## VI. Conclusion

This paper has discussed a simple complete analytical modeling and simulation of a new radiation calorimeter operating in the range of -50 °C to 150 °C. Within this model the thermal properties of the calorimeter components materials have been taken as functions of temperature. The theoretical modeling of heat exchange has taken into account the three modes of heat transfer. It has been seen that theoretical change in the calorimeter surrounding temperature leads to heat variation on the measuring cell to 2.9 mW.K$^{-1}$ when no heating power was generated inside. Then, it is preferable to work in environment with fixed temperature for limiting thermal disturbance on the cell. It has also been observed that the measuring cell is relatively sensitive to the heating power provided to it by the regulation. The calorimeter measuring principle is based on heat radiation, the DUT support and current leads limit the radiation. The simulation predicts that the ratio of the radiated heat power on the total heat generated by the DUT varies from 99.8 % to 99.95 % when its leads supplies are in constantan material. These values are obtained with high surfaces emissivity of the DUT and the measuring cell. The Nextel® velvet coating 811-21 is an interesting solution for that problem. The calculations performed in thermal transient regime to appreciate the thermal dynamic of the measuring cell showed however that this latter is slow. Indeed for the initial temperature of 17 °C the cell needs 55 hours to reach a temperature of -20 °C, when its support is in glass fiber epoxy composite material in vacuum. In practice the solution can be to encourage thermal convection with nitrogen gas inside the calorimeter before its definitive evacuation or the utilization of cell support material with better thermal conductivity [2]. Finally our model showed that a heating power of 0.1 mW dissipated by the DUT is measurable in theory by instruments having a resolution of 1 mK. In practice we obtained an accuracy of ± 5 % for 13.7 mW heat power dissipated for temperature ranging from -20 °C to 75 °C. The comparison between the model and experiment of the thermal dynamic response of the cell during cooling shows that the temperature gap between them remains in the worst case lower than 5 °C for the cooling time going from 0 to 55 h. However the gap in the measurable heat power seen between the theory and the experiment has been attributed to the transient thermal regime of the cell that does not reach its steady state regime. From these analytical results the calorimeter has been successfully implemented and patented.